\documentclass[useAMS,usenatbib]{mn2e}

\title[A Super-Earth  in a trap]
{A Super-Earth caught in a trap}
\author[E. Podlewska and E. Szuszkiewicz]{E. Podlewska$$\thanks{E-mail:
edytap@univ.szczecin.pl (EP)} and E. Szuszkiewicz$$\thanks{E-mail: 
szusz@univ.szczecin.pl (ES)} \\
Institute of Physics and CASA*, University of Szczecin, ul. Wielkopolska 15,
70-451 Szczecin, Poland}
\begin{document}

\date{}

\pagerange{\pageref{firstpage}--\pageref{lastpage}} \pubyear{}

\maketitle

\label{firstpage}

\begin{abstract}
This paper is an extension of the work done by \citet{pierens}
in which they have investigated the behaviour of a two-planet 
system embedded in a protoplanetary disc. They have put a 
Jupiter mass gas giant on the internal orbit and a lower mass 
planet on the external one. We consider here a similar problem 
taking into account a gas giant with masses in the range of $0.5$ 
to $1M_J$ and a Super-Earth (i.e. a planet with  mass $\leq 10
M_\oplus$) as the outermost planet. By changing disc parameters
and planet masses we have succeeded in getting the convergent 
migration of the planets which allows for the possibility of 
their resonant locking. However, in the case in which the gas 
giant has the mass of Jupiter, before any mean motion first order
commensurability could be achieved, the Super-Earth is caught in 
a trap when it is very close to the edge of the gap opened by the 
giant planet. This confirms the result obtained by \citet{pierens} 
in their simulations. Additionally, we have found that, in a very 
thin disc, an apsidal resonance is observed in the system if the 
Super-Earth is captured in the trap.
Moreover, the eccentricity of the small planet remains low, while 
the eccentricity of the gas giant increases slightly due to the 
imbalance between Lindblad and corotational resonances. We have
also extended the work of \citet{pierens} by studying analogous
systems in which the gas giant is allowed to take Sub-Jupiter 
masses. In this case, after performing an extensive survey over
all possible parameters, we have succeeded in getting the 1:2 
mean motion resonant configuration only in a disc with low aspect 
ratio and low surface density. However, the resonance is maintained
just for few thousand orbits. Thus, we conclude that for typical 
protoplanetary discs the mean motion commensurabilities are rare
if the Super-Earth is located on the external orbit
relative to the gas giant.
\end{abstract}

\section{Introduction}

Among over 300 extrasolar planets discovered so far only a few
Super-Earths, that is planets with masses in the range of 2-10 $M_{\oplus}$, 
have been observed.
However, more candidates are just waiting to be confirmed and new 
discoveries are going to be announced, so we can soon find ourselves
in the middle of the Super-Earth epoch. Some of such low-mass planets 
might exist in the neighborhood of gas giants and by means of 
numerical simulations it is possible to predict what are the most
common configurations of extrasolar systems with Super-Earths
and what will be detected by present and future
observational programmes.
It is believed that the observed architecture of the planetary systems might
be an outcome of the large scale orbital migration induced by the
disc-planet interactions. Depending on the planet masses and on the disc
properties, two main regimes of orbital migration can be distinguished
\citep{ward97}. 

The migration time for low-mass planets or planetary embryos embedded
in a gaseous disc has been derived first by summing the
differential torques induced by the Lindblad resonances 
\citep{goltrem1979, ward97} and
improved later by taking into account
the contribution from the corotation torque \citep{tanaka02}
\begin{eqnarray}
\tau_{I}=(2.7+1.1 \gamma)^{-1} \frac{M_*}{m_p}\frac{M_*}{\Sigma
{r_p}^2} \left( \frac{c}{r_p \Omega_p}\right)^2 {\Omega_p}^{-1}
\label{tanaka}
\end{eqnarray}
Here $m_p$ is mass of the planet,
$r_p$ is the distance from the central star $M_*$, $\Sigma$ is the disc
surface density, $c$ and $\Omega_p$ are respectively
the local sound speed and the angular
velocity. The coefficient $\gamma$ depends on the disc surface
density profile, which is expressed as $\Sigma(r) \propto r^{-\gamma}$.
Eq.~(\ref{tanaka}) refers to type I migration, where the disc
response is linear and the planet is not massive enough to
perturb significantly 
the mass distribution of the gas in the disc.
However, recent studies showed a strong departure from the linear theory. 
It has been found that in non-isothermal discs with high opacity
\citep{paarmel06} in
the presence of an entropy gradient  \citep{paapap08}
the sign of the total torque can change reversing in this way the
direction of the migration.

For high-mass planets the disc response becomes non linear and a gap forms
in the disc around the planet orbit. If the gap is very
clean
and the disc is stationary, the evolution of the planet
 is determined by the
 radial velocity drift in the disc  \citep{linpap86}, namely
\begin{eqnarray}
v_r=\frac{3\nu}{2r_p}.               \label{vmigr}
\end{eqnarray}
The migration time of the planet can be estimated as
\citep{linpap93}
\begin{eqnarray}
\tau_{II}=\frac{2 {r_p}^2}{3 \nu}.        \label{kl}
\end{eqnarray}
However, recent numerical simulations \citep{edgar} showed that
the migration 
rate depends also on the
planet
mass and on the disc surface density profile, which can be written as
\begin{eqnarray}
\tau_{II}=\frac{m_p}{3 \nu \Sigma (r)}.      \label{redgar}
\end{eqnarray}
The latter estimation gives a dependence on the surface density
which is in good agreement with
hydrodynamic
simulations, although the  variation with  the
disc viscosity appears to be much weaker than expected from  analytical 
predictions \citep{edgar,edgar2008}.

For intermediate-mass planets which open the gap only partially,
type III migration has been proposed \citep{maspap}.
This type of migration occurs if the disc mass is much higher than the mass
 of the planet.

It is than required to employ numerical simulations
in order to determine how the relative migration rate of the planets
varies in details with their masses and with the parameters specifying
the properties of the disc.

Different rates of planetary migration can lead to the resonant
capture of the planets as it was shown for two giant planets
in \citet{kley} or \citet{kbp04}.
Also low-mass planets may
undergo convergent
migration and form a resonant structure \citep{papszusz}.
In a previous paper of ours \citep{paperI}, we have considered a system with
low- and high-mass planets
(a Super-Earth on the internal orbit and a Jupiter-like planet on the 
external one) and we have found that they are captured into 3:2 or 4:3 
mean motion resonances if the disc properties allow for convergent 
migration. 
Without drawing a too close analogy to the formation of the Solar 
System, it has been pointed out that the Hilda and Thule groups of
asteroids are  
in the interior 3:2 and 4:3 resonances with Jupiter. Similar 
configurations with a Super-Earth instead of an asteroid might be present
in extrasolar planetary systems. In order to draw such a conclusion,
a long-term stability analysis of the obtained resonant structures should
be performed. 

In the present work we examine a slightly different situation. 
As in \citet{paperI} planets form a close pair and are embedded in the 
gaseous disc, but
this time
the Super-Earth lies on the external orbit. 
This case has been already studied by  \citet{thommes05}
 and more recently by \citet{pierens}
(hereafter PN08).
\citet{thommes05}
has found that a Jupiter mass planet
can act, at least for a certain time, as a safety net for low-mass planets,
capturing them into orbital commensurabilities.
In his calculations, the low-mass planet, which is allowed to accrete
the mass, migrates towards the Jupiter and ends up in a 1:2 or 2:3
resonance depending on the surface density of the disc.
Contrary to his result, PN08
have noticed that  a low-mass planet (in the range of 3.5 - 20 $M_{\oplus}$)
can be  trapped at the outer edge of the gap
opened by the gas giant. Planets cannot therefore get close
enough as it is necessary to attain
a mean motion resonance. 
Such planet trapping mechanism 
is possible at the steep and positive surface density
gradient in the radial profile of the gaseous disc where, the 
corotation torque compensates the differential Lindblad torque 
\citep{masset}. 
More recently \citet{paandpa}
have studied the nonlinear effects arising
in the coorbital region of planets of a few Earth masses.
They have found that any  positive density gradient in the disc
can act as a protoplanetary trap.

The investigations of \citet{thommes05} have been performed  
 using a hybrid code, that
combines an N-body component with a one-dimensional viscous disc model.
In this way the effects of Lindblad torques are
well reproduced, but the corotation torques
acting on the low-mass planet are not taken into account. Employing  
a 2D hydrodynamical code, PN08 have been able to include
properly both types of torques.
It has been shown by PN08 that the initially convergent migration of a 
Jupiter mass gas giant and an outer planet less massive than 20~$M_{\oplus}$
stops when the low-mass planet approaches the outer edge of the gap
formed by the gas giant. Later on, the low-mass planets migrate outward.
In the present work we confirm the results of PN08 in the case of a
Super-Earth and a Jupiter-like planet. The only difference is that in the 
calculations of PN08 the low-mass planet was able to accrete matter from 
the disc and if its mass exceeded the value of 20  $M_\oplus$, the planets 
ended up in a 
mean-motion resonance. In this paper we are interested in Super-Earths,
so we did not take  into account accretion. 

As an extension of the work of PN08, 
we have investigated the evolution of a Super-Earth also in the presence of
Sub-Jupiter  gas giants, whose  masses are lower than that of Jupiter.
For typical disc properties the gap opened by the gas giant
is very wide and the positions of all first order
mean motion commensurabilities are located inside the region affected by the
gap.
If the gas giant  is assumed to be less massive than Jupiter (e.g. $0.5
M_J$), the gap opened by the planet is narrower,
which gives a chance for attaining a first order mean motion resonance.
However, in this situation the migration of the gas giant is faster than
that of the low-mass planet
causing the divergent relative motion of both planets. For this reason,
in order to get the commensurability one needs to slow down the Sub-Jupiter
or to speed up the Super-Earth. Finally,
we have achieved the 1:2 mean motion resonance  
for very thin discs and low surface density. This configuration, however,
has not been maintained till the end of the simulations.

This paper is organized as follows. In Section 2 we describe our
numerical set-up. In Section 3 we present the procedure used
in order to achieve
the convergent migration of a Super-Earth and a Jupiter-like
planet. Section 4 is 
dedicated to the results obtained in the evolution of such a system of
planets.
We also discuss in this Section
the eccentricity behaviour together with the 
analysis of the sensitivity of our results to numerical parameters
and the occurrence of the apsidal resonance.
In Section 5 we investigate the scenario
in which the Jupiter-like planet is substituted by a gas giant with
a smaller mass.  
Finally, we summarize and discuss our findings
in Section 6.

\section{Description of the numerical simulations}

\begin{figure}
\vskip 6.5cm
\includegraphics{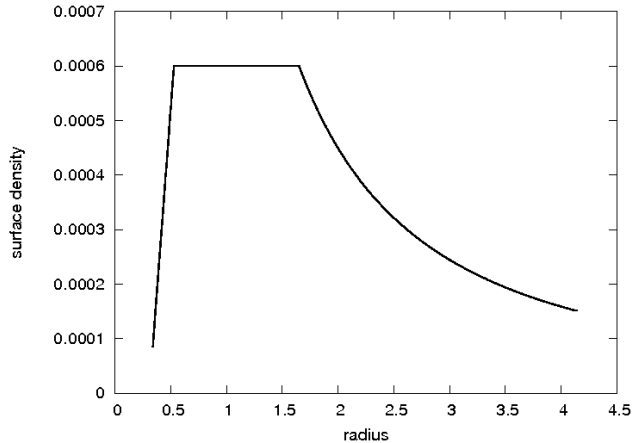}
\caption{\label{fig1}{The initial surface density profile of the disc. The
planets were initially located in the flat part of this profile.
}}
\end{figure}

We have performed numerical simulations of a system containing
two interacting planets and a gaseous protoplanetary disc with which the
planets interact. One of the planets is a Super-Earth and the other is
a gas giant. They are considered as point masses orbiting
around the central star with mass $M_*=M_\odot$.
The disc undergoes near-Keplerian rotation and its vertical 
semi-thickness $H$ is small in comparison with 
the distance $r$ from the central star. We assume a constant aspect
ratio $h=H/r$, so that the temperature profile of the disc is $T
\propto r^{-1}$. 
The best choice of  coordinates for this problem is that of
cylindrical coordinates ($r$, $\varphi$, $z$) with the
origin located at the position of the central star.
The equations of motion can be vertically 
averaged. In this way the problem is reduced to two dimensions,
given by the radial and 
azimuthal directions. 
The evolution of the gaseous disc in our system is governed by the
continuity equation and two equations of motion (for the full formulation,
see \citet{nelson2000}). We are not solving the full energy equation
because of the high computational cost.  
The locally isothermal equation of state of the gas in the disc is adopted
instead.  
However, it has been noticed that
solving the energy equation could have dramatic effects on the migration
of the low-mass planet \citep{paarmel06}. Namely, in non-isothermal
discs the 
migration can be directed outward.
We have used the same Eulerian hydrodynamic code NIRVANA \citep{ziegler} 
as in our previous paper \citep{paperI}. The details of the numerical scheme 
can be found in \citet{nelson2000}.
We have checked that the results obtained with our version of
the NIRVANA code are in very good agreement with those obtained in the
framework of the project "The Origin of Planetary Systems" 
\citep{devalborro}. 
Following a common practice, in order to assess
 the robustness of our results we 
have repeated part of  our investigation with a second code,
which in our case was FARGO \citep{masset2000}.
The initial surface density profile 
of the disc $\Sigma(r)$ is taken to be flat at the planet location.
Its overall  shape shown in (Fig.~\ref{fig1})
has been constructed for computational convenience, namely to avoid an
unwanted behaviour of the disc at the boundaries of the computational domain. 

The surface density value $\Sigma_0=6\times10^{-4} 
\left(\frac{M_\odot}{(5.2AU)^2}\right)$ of the flat part of this profile
corresponds to the minimum mass solar nebula (MMSN) which 
consists of a gas weighing two Jupiter masses spread out within a
circular area of radius equal to the mean distance of Jupiter from the Sun.
The adopted unit of length corresponds to $5.2 AU$.
The unit of time is given by $(GM_*/{r_p}^3)^{-1/2}$ ($G$ is the
gravitational 
constant, $M_*$ denotes the mass of the star and $r_p$ the initial radial
position of the 
inner planet). This quantity amounts to ($1/2\pi$) times the
orbital period of the initial orbit of the inner planet and it will be
called just an "orbit" throughout the whole paper.
The initial locations of the planets and their masses are given in 
Table \ref{tab1}.
At the beginning of all runs we have put both planets on circular orbits 
in the flat part of the surface density profile. 
 The computational domain extends between $r_{min}=0.33$ 
and $r_{max}=4.15$ in the
radial direction as it is  shown in Fig. \ref{fig1}.
We have also performed one
run with $r_{max}=5$ in order to check the evolution for a larger initial
separation of the planets.
The azimuthal angle $\varphi$ takes its values in the interval
$[0,2\pi]$.
The disc is divided into $400\times 512$ grid cells in the radial and 
azimuthal directions respectively.
For the model with $r_{max}=5$ we increased the resolution in such a
way
that 
the size of each 
single grid cell remains unchanged. 
 We have also run one
simulation with a higher number of grid
cells
($576\times 986$)
and we have found no significant difference in the migration rate and the 
eccentricity evolution, so we assume that our standard resolution is
appropriate
 for our experiments.
The radial boundary conditions were taken to be open, so that
the material in the
disc can outflow through the boundaries of the computational domain.
In this case the profile of the disc
changes with  time. In order to avoid  unrealistic effects due to
gas depletion, we have
run our simulations for no longer than $10^4$ orbits.
The potential is softened with softening parameter $\varepsilon=0.8H$. 
Some of the simulations have been run also with
$\varepsilon=0.6H$ and $\varepsilon=1.3H$ in order to check the 
influence of the softening length
on the evolution of the planets.
In the calculation of the gravitational potential 
of the planets
we do not exclude the matter
contained in the planet Hill sphere.
The selfgravity of the disc is not taken into account.
In order to get the convergent migration of the planets when the gas giant is on
the internal orbit and the Super-Earth on the external one, we have performed
a series of numerical simulations using different disc parameters. All
simulations are summarized in Table~\ref{tab1}. We have run our
experiments with 
constant aspect ratio $h$ ranging from $0.03$ to $0.05$ and constant
kinematic 
viscosity $\nu=2 \cdot 10^{-6}$ in dimensionless units (this
corresponds to 
the $\alpha$ parameter being equal to $2.2 \cdot 10^{-3}$
for $h=0.03$ and $8 \cdot 10^{-4}$ for $h=0.05$
). In one
case,
a
higher viscosity value  ($\nu=5 \cdot 10^{-6}$) has been used
in order to obtain that the
gap opened by the gas giant is very narrow. The last column in
Table~\ref{tab1} shows 
whether 
we have achieved convergent migration or not. In this table
"convergent/divergent"
means that the relative migration of the planets reversed during the evolution.
\begin{table*}
\centering
\caption{\label{tab1}{In this table we summarize all the  performed 
simulations. 
The first column
shows the model number, the second the initial semi-major axis of the
Jupiter, the third the initial semi-major axis of the Super-Earth,
the fourth the outer edge of the disc, the fifth the mass of the
Super-Earth, the sixth the mass of the gas giant, the seventh
the surface density, the eighth the aspect
ratio, the 
ninth the kinematic viscosity and the tenth the kind of
relative migration of the planets.
The comment "convergent/divergent"
means that relative migration of the planets reversed during the evolution.
}}
\begin{tabular}{@{}clcccccccc@{}}
\hline
model&$r_{p1}$& $r_{p2}$& $r_{max}$ & mass of the& 
 mass of the&surface density& h & $\nu$ &migration   \\
 & & & & Super-Earth ($M_\oplus$)& gas giant ($M_J$)& & \\
\hline
1 & 1    & 1.62   & 4.15 & 5.5&1 &$ \Sigma_0$ &0.05& $5\cdot 10^{-6}$
 & divergent\\
 \hline
2 & 1    & 1.62   & 4.15 & 5.5&1 &$ \Sigma_0$ &0.05& $2\cdot 10^{-6}$
 & divergent\\
 \hline
3 & 1    & 1.35  &  4.15 & 5.5 & 1& $\Sigma_0$ &0.03& $2\cdot 10^{-6}$
&convergent/divergent\\
 \hline
4 & 1    &1.62   &  4.15 & 5.5& 1&  $\Sigma_0$ &0.03& $2\cdot 10^{-6}$
&convergent/divergent \\
 \hline
5 & 1    & 1.58   &  4.15 & 5.5 & 1&  $\Sigma_0$ &0.03& $2\cdot 10^{-6}$
&convergent/divergent\\
 \hline
6 & 1    & 2.2   & 5 & 5.5 & 1& $\Sigma_0$ &0.03& $2\cdot 10^{-6}$
&convergent/divergent\\
 \hline
7 & 1    & 1.62   & 4.15 & 5.5&1 &$2  \Sigma_0$ &0.05& $2\cdot 10^{-6}$
 & convergent/divergent\\
 \hline
8 & 1    & 1.62   & 4.15 & 10& 0.5 &$1.5 \Sigma_0$ &0.05& $2\cdot 10^{-6}$
&divergent\\
 \hline
9 & 1    & 1.62   & 4.15 & 10& 0.5 &$2.5 \Sigma_0$ &0.05& $2\cdot 10^{-6}$ 
&divergent\\
 \hline
10 & 1    & 1.62   & 4.15 & 10& 0.6 &$1.5 \Sigma_0$ &0.05& $2\cdot 10^{-6}$ 
&divergent\\
 \hline
11 & 1    & 1.62   & 4.15 & 10& 0.7 &$1.5 \Sigma_0$ &0.05& $2\cdot 10^{-6}$ 
&divergent\\
 \hline
12 & 1    & 1.62   & 4.15 & 10& 0.5 &$\Sigma_0$ &0.03& $2\cdot 10^{-6}$
&convergent/divergent\\
 \hline
13 & 1    & 1.62   & 4.15 & 10& 0.5 &$\Sigma_0$ &0.04& $2\cdot 10^{-6}$
&convergent/divergent\\
 \hline
14 & 1    & 1.62   & 4.15 & 10& 0.5 &$0.5 \Sigma_0$ &0.03& $2\cdot 10^{-6}$
&convergent\\
 \hline
\end{tabular}
\end{table*}

\section{Looking for the convergent migration of a Super-Earth and a Jupiter}

Similarly as in our previous paper, we have tried to determine the initial
conditions appropriate for the convergent migration using the analytic formulae 
given by equations (\ref{tanaka}) for the Super-Earth and (\ref{kl}) or
(\ref{redgar}) for the gas giant. However, we have
found as in \citet{edgar} that in an isothermal viscous
disc a giant planet  does not obey the standard 
type II migration and hence migration does not proceed on the viscous
timescale. That is why we have performed a series of numerical experiments
in order to examine under which conditions it is possible to obtain
convergent migration. 

Using the full 2D hydrodynamical code NIRVANA
we have calculated the evolution
of the orbital elements
of the planets embedded in the gaseous disc in the region where the exterior
first order mean motion resonances are located. The gas giant planet has
been placed always at the distance 1 in our units and the Super-Earth 
further away from the star
(see Table \ref{tab1}), starting
from a small separation between planets of the order of a few Jupiter's
Hill radii, and then increasing it.
Planets are not allowed to accrete matter and their masses are fixed 
for the whole time of the calculations.

\begin{figure}
\vskip 6.5cm
\includegraphics{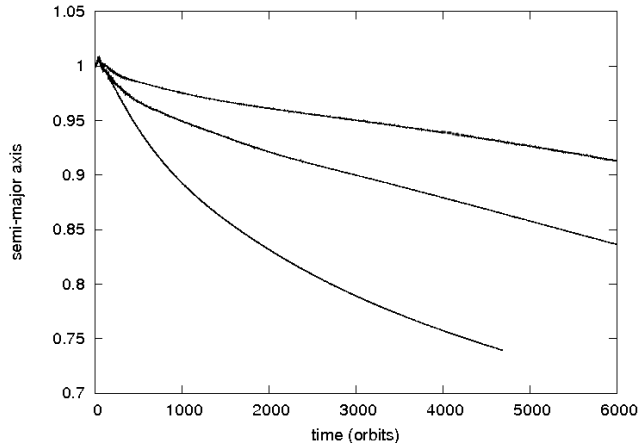}
\caption{\label{modrho}{The semi-major axis of the Jupiter for models 1
(lowest curve), model 2 (middle curve) and model 3 (upper curve) of
Table \ref{tab1}.
}}
\end{figure}

We have analyzed different disc models changing the aspect ratio,
the kinematic viscosity and the surface density of the disc.
These investigations allowed us to find
the conditions 
under which the migration 
is convergent at the beginning of the evolution. This does not imply that
it will remain convergent throughout the whole run. In fact, in most of our
models divergent migration is the final outcome.
 If the kinematic viscosity is
high, for instance $\nu=5\cdot 10^{-6}$ ($h=0.05$ and
$\Sigma=\Sigma_0$),  the 
migration of the Jupiter is fast and the relative
migration of both planets is divergent (Table~\ref{tab1}, model 1).
For gas giants the migration speed is evaluated to be proportional to 
the viscosity of
the disc \citep{linpap93}, so if we take a  lower value of $\nu$, the
migration  
of the Jupiter is slower. Assuming for the kinematic viscosity the value
$\nu=2 \cdot 10^{-6}$ without changing other parameters (Table
\ref{tab1}, model 2), we have obtained indeed a slower migration
than in model 1 as it
is shown in  Fig. \ref{modrho}, but still the relative motion of both planets
is divergent. 
We have noted that the migration of the gas giant is
faster than that of the 
Super-Earth even after pushing the kinematic viscosity as low as
$\nu=8 \cdot 10^{-7}$. We conclude that convergent migration cannot be
achieved changing
 only the viscosity of the disc.
It was shown by \citep{edgar} that the migration
of the gas giant depends
also on the surface density, 
so we have continued our search for convergent  migration changing
other disc 
properties.
A full analysis of the
influence of the disc parameters on the gap profile and thus on the migration of
the planet can be found in \citet{crida}. 
We have performed
simulations with relatively low $\nu$ ($2 \cdot 10^{-6}$) changing
$h$ and $\Sigma$.
First, we have set $\Sigma =\Sigma_0$ and achieved
the convergent migration of both planets for a
very thin disc with aspect ratio $h=0.03$
(for example model 4). In this case, the
gap opened by the gas giant is very wide and the migration of the Jupiter
is slower than that of the Super-Earth, which causes that
 the radial distance between the planets decreases. 
Convergent migration has been obtained also keeping the aspect ratio fixed
($h=0.05$) and changing the surface density. Taking 
$\Sigma=2\Sigma_0$,
(as in model 7) we have got 
a migration of the Super-Earth which is faster than in model 2 in
accordance to the simple 
approximation of type I migration
\citep{tanaka02} and at the same time the migration is convergent.

\section{The evolution of the Super-Earth and the Jupiter}

\subsection{The semi-major axis evolution}\label{section:c}

In PN08 it has been discussed the evolution of a two-planet system
containing a low-mass planet with a mass in the range of
$3.5-20M_{\oplus}$ and a Jupiter mass planet, both embedded in a disc
with the surface density  $\Sigma \propto r^{-3/2}$. According to PN08
this evolution should end up with the trapping of the low-mass planet
near the outer edge of the gap produced by the Jupiter mass planet.
In this Section we are interested in the  
possibility of resonant locking of a
Super-Earth with mass $5.5M_{\oplus}$ and a Jupiter mass planet
embedded in a disc with a flat surface density profile.
In order to check whether the outcome of the evolution of the system
considered here is the same as in PN08, we have performed a series of
simulations which are described below. 

\begin{figure}
\vskip 6.5cm
\includegraphics{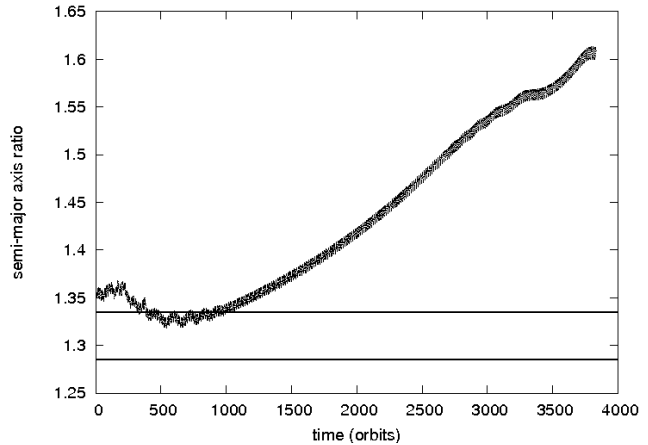}
\caption{\label{fig2}{The semi-major axis ratio of the planets versus time.
The initial planet separation is 0.35 (model 3 of Table \ref{tab1}).
The planets have
been captured in the 2:3 mean-motion resonance (exact position of the
commensurability is 1.31) but the resonant structure did not last and
the migration became divergent. The horizontal lines denote the width
of the 2:3 commensurability.
}}
\end{figure}

In Fig.~\ref{fig2} we show the evolution of the semi-major axis ratio of
the planets in the case where the initial planetary orbital separation is
0.35 (Table \ref{tab1}, model 3). 
The semi-major axis ratio has been 
calculated
by dividing the semi-major axis of the Super-Earth by the semi-major axis
of the gas giant. The convergent migration brings planets into
the 2:3 mean motion commensurability. The horizontal lines in Fig. \ref{fig2}
show the width of this resonance \citep{lecar}. However, this configuration 
lasts just
for a few hundreds of orbits and after that  the relative distance between 
the planets increases. 
The Super-Earth migrates outward 
and the Jupiter migrates  inward, as
expected,
with a rate determined by its mass and by the disc parameters.
It is then clear that the evolution of the Super-Earth is responsible 
for reversing 
the convergent relative motion of the planets into
a divergent one after roughly 1000 orbits.

\begin{figure*}
\begin{minipage}{160mm}
\vskip 6.5cm
\includegraphics{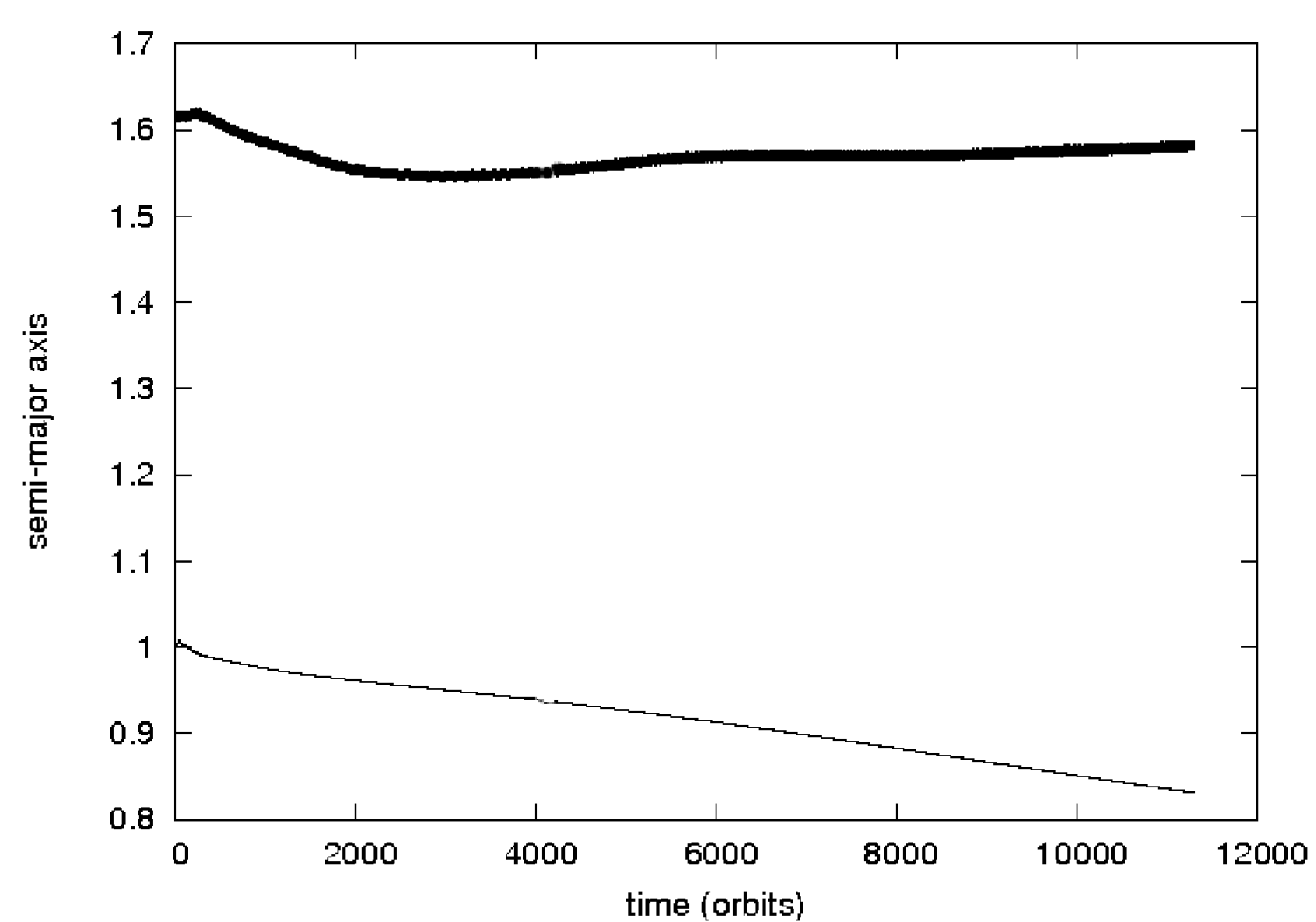}
\includegraphics{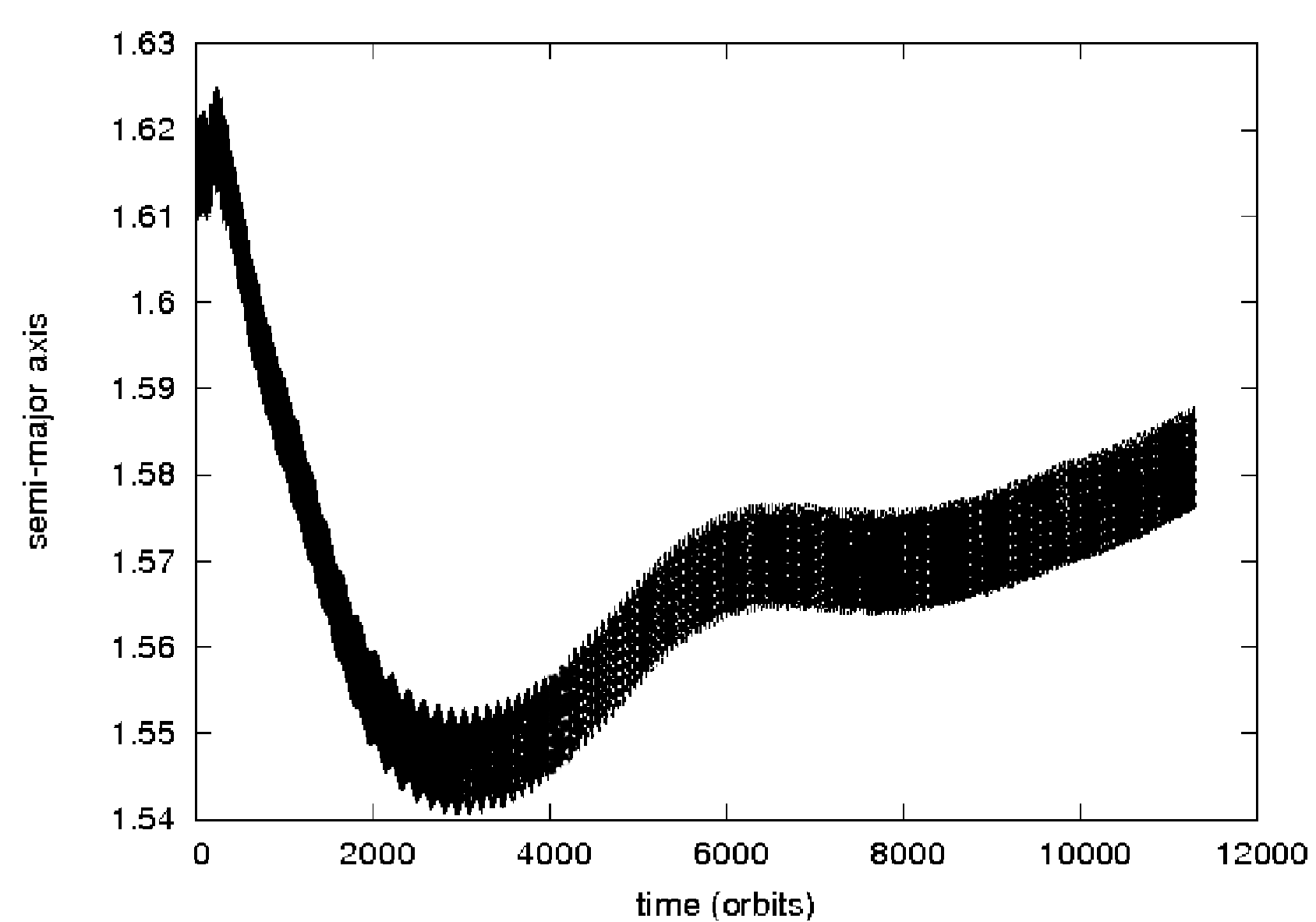}
\caption{\label{fig4}{Left panel: The semi-major axis of planets with
initial separation of 0.62 (model 4 of Table \ref{tab1}).
The upper curve refers to the Super-Earth and the lower one to the gas
giant.
Right panel: The evolution of the semi-major  axis of the Super-Earth
for the same model, but with a different scale in the vertical
axis. At the beginning the Super-Earth
migrates inward but later on the migration has changed the direction because
the planet is trapped and it moves together with the edge of the gap.
}}
\end{minipage}
\end{figure*}

\begin{figure}
\vskip 6.5cm
\includegraphics{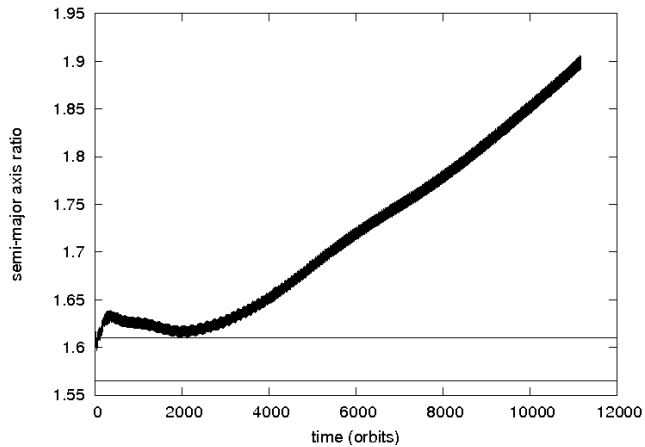}
\caption{\label{fig5}{The semi-major axis ratio of the planets with
the initial
separation of 0.62 (model 4 of Table \ref{tab1}).
The exact position of the resonance is 1.58. The migration becomes divergent
before the planets could enter the resonant region denoted by the horizontal
lines.
}}
\end{figure}

Next, we have checked the possibility
of the occurrence of the other first order commensurability found by
\citet{thommes05}, namely 1:2.
To this aim, we have placed the Super-Earth further out from the Jupiter,
at the relative orbital separation of 0.62 (Table \ref{tab1}, model 4). 
The evolution of this configuration
is shown in Fig.~\ref{fig4} (left panel). 
The convergent migration continues 
for about 2000 orbits and after that, similarly as in model 3, the Super-Earth 
migrates outward. In the right panel of Fig. \ref{fig4}
we have plotted again the behaviour of 
the Super-Earth, but this time using a different scale, in order to see its 
orbital evolution in full detail. 
In Fig.~\ref{fig5} we have illustrated 
the evolution of the ratio of the semi-major axes of the planets. This
ratio lies outside the resonance region whose width is  
marked by two horizontal lines \citep{lecar}. 
Next, we have put the 
planets exactly in the 1:2 resonance (Table \ref{tab1}, model 5) and we have
obtained the same result seen in model 4. 
For a disc with higher surface density and aspect ratio $h=0.05$
 (as in the model 7), we have also
obtained convergent migration at the beginning, but later on the
relative migration of the planets reversed as in the
previous models. As a result, no
mean motion commensurability has been attained.    
 Our conclusion is that even the most 
distant first
order 1:2 mean motion resonance cannot be achieved in this way.

\begin{figure}
\vskip 6.5cm
\includegraphics{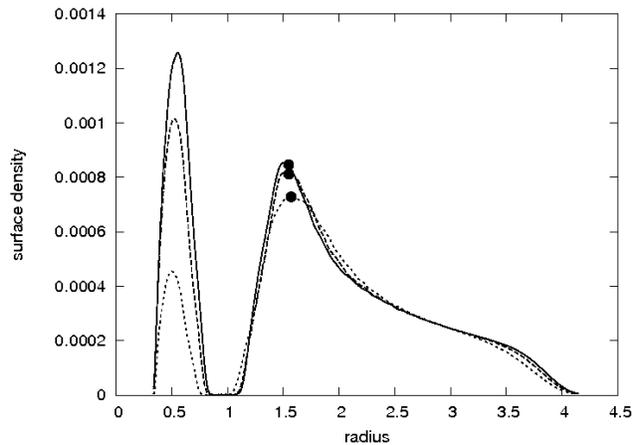}
\caption{\label{fig6}{The surface density profile of the disc after 2355
(solid line), 3925 (dashed line) and 7850 (dotted line) orbits
for model 4 of Table \ref{tab1}.
 The dots
denote the positions of the Super-Earth.
}}
\end{figure}

Similar results have been presented by PN08 in the case in which the
outer planet has
mass 
$10M_\oplus$ or $20 M_\oplus$. The answer to the question why the
Super-Earth is not able to form resonant configuration with the gas giant is
illustrated
in Fig.~\ref{fig6}, where 
we show the surface density profile changes in model 4 
after about 2355, 3925 and 7850 
orbits, together with the position of the  Super-Earth marked 
by black dots. 
It can be seen 
that the outer edge of the gap in moving slowly outward but the planet remains 
trapped in the high surface density region at the edge of the cavity 
till the end of the simulation.
The same situation was observed in PN08. However, in their paper the outer
planet is more massive than in our simulations and it is stopped closer to
the gas giant where the vortensity gradient is larger.
From the analysis of the  position of the Super-Earth relative to the gap,
we conclude that the outward migration of this planet seen
in Fig. \ref{fig4}
is caused by the evolution of the gap opened by the gas giant in the disc.
So the
migration of the two planets is convergent until the Super-Earth
is trapped. After that, the migration reverses and the Super-Earth
moves together with 
the expanding edge of the gap 
as reported in PN08.

The models discussed till now have one common feature, namely the 
planets   are placed initially very close to the resonance (models 3, 4 and
7) 
or exactly in the resonance (model 5). This choice has been made 
for computational reasons. However, locating the planets far away from 
the first order
commensurabilities would be useful for the interpretation of our results.
Having this in mind,
we have performed also one experiment with a relatively large initial orbital 
separation
between planets 
like in PN08
 (Table \ref{tab1}, model 6). The outcome of the
evolution of this model is that   the Super-Earth is caught again in the
trap at the edge of the gap.

\begin{figure}
\vskip 6.5cm
\includegraphics{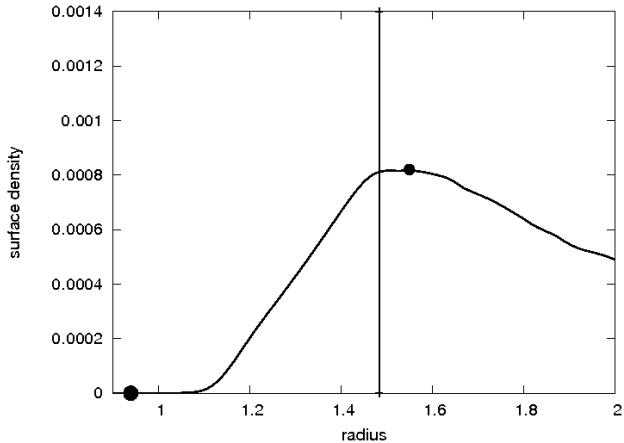}
\caption{\label{resgap}{The surface density profile of the outer edge of the
gap in the model 4 after 3925 orbits. The dots denote the positions of
the planets, the vertical line the
location of the 1:2 mean motion resonance. The Super-Earth is captured in
the trap so it cannot approach the commensurability.
}}
\end{figure}

In all the models where we have obtained  convergent migration the gap is
wide and the position of the 1:2 resonance is always inside the gap.
This fact is illustrated in  Fig. \ref{resgap}, where we plot the
gap profile 
of the outer part of the disc for  model 4 after 3925 orbits. 
The dots denote the position of the Super-Earth and of the Jupiter. 
The vertical line shows the location of the 1:2 commensurability.  
The Super-Earth is captured in the trap which prevents the further
migration towards the resonance.

In order to avoid the process connected to the cavity opening which can 
influence the disc structure in the vicinity of the gas giant, we have 
performed a simulation
in which the Jupiter was initially located in a fully formed gap. 
We have used in this simulation a disc profile in which the shape of
the gap has been obtained from that of model 4 after eliminating the
bumps at the edges of the cavity. In this way the surface density
maxima outside the edges, which are a consequence of the gap formation
and could affect the results, are absent.
Also in this case,  the
 Super-Earth migrated slowly inward and was captured in the trap as in
the previous models.
The assumption of the existence
of the initial gap enabled us to study the evolution in
a more
realistic way.

\subsection{The eccentricity evolution}
\begin{figure}
\vskip 6.5cm
\includegraphics{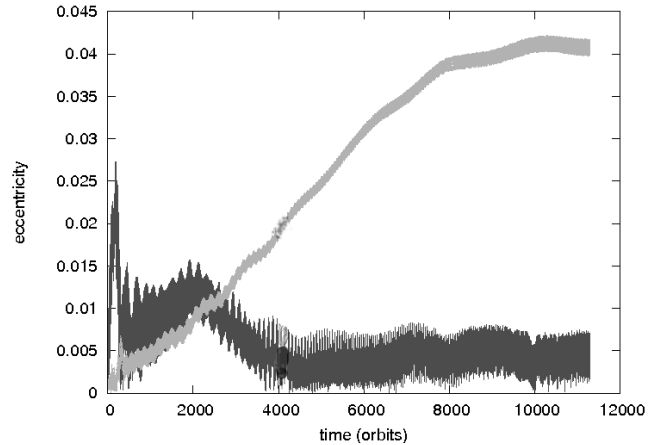}
\caption{\label{eccentric}{The eccentricity evolution of the gas giant (gray
colour)
and the Super-Earth (black colour) for model 4 of Table \ref{tab1}.
}}
\end{figure}

We  investigated also the evolution of the eccentricities of the planets. 
At the beginning of our simulations both planets are located on 
circular orbits. 
First, let us have a look
at the eccentricity of the Super-Earth. 
When  the  migration  is convergent, the planets 
approach the 1:2 resonance 
and the Super-Earth's eccentricity increases  slightly,
see for instance Fig. \ref{eccentric}, which describes the case of
model 4 of Table~\ref{tab1}.
As soon as the Super-Earth is trapped at the edge of the gap, the 
planets depart from the resonance and 
the Super-Earth's eccentricity is damped 
 by the tidal interaction with the disc 
to a low value and remains at this level till the end of the simulation. 
In case where the migration is always divergent since the beginning of the
simulation,
the eccentricity of the Super-Earth remains low during the whole time of
the evolution.

\begin{figure}
\vskip 6.5cm
\includegraphics{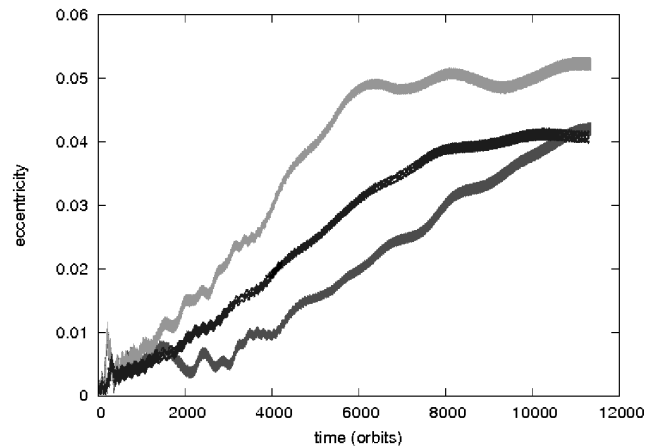}
\caption{\label{fig7}{The eccentricity of the gas giant with
$\varepsilon=0.6H$ (light gray curve), $\varepsilon=0.8H$ (dark gray
    curve) and
$\varepsilon=1.3H$ (black  curve) for  model 4 of  Table \ref{tab1}.
The eccentricity increased in all cases.
}}
\end{figure}

As far as the gas giant is concerned,
during its evolution in the disc 
the eccentricity is very low in most cases. 
However, we have found an
interesting feature in the discs with  aspect ratio $h=0.03$. Namely, the
eccentricity
increases for the first few thousand orbits, later on 
its growth is slowed down and at the end of the simulation approaches 
the value of 0.04-0.05. We have investigated this issue in detail and we have
checked the dependence of the eccentricity growth on numerical
parameters such as the softening length and the  grid resolution.
We have performed simulations with the same set-up of model 4 using for
the
softening parameter the values $\varepsilon=0.6H$, $\varepsilon=0.8H$ and
$\varepsilon=1.3H$, which correspond respectively to the following values
of the radius
of the Jupiter's Hill sphere: $0.26$, $0.35$ and $0.58$. In
Fig. \ref{fig7} we compare the 
evolution of the eccentricity of the gas giant with these three parameters.
Despite some differences at the initial stages of the
evolution, the result is approximately the same in each simulation.
It is important to mention at this point that the evolution of the gas giant
eccentricity is the same in all calculated models with $h=0.03$.
In order to check the effects of the grid resolution
we have repeated the run of
model 4 with lower resolution ($192 \times 256$ grid cells in the radial and
azimuthal direction respectively) and we 
didn't observe any eccentricity
growth. This  suggests that to detect this behaviour one needs
higher grid resolutions.
 We have also run this
simulation with a higher number of grid
cells
($576\times 986$)
and we have found no significant difference in the migration rate and the
eccentricity evolution. 

An extensive  discussion is going on in the literature on the 
possible mechanisms for exciting eccentricities
(e.g \cite{arty91}, \cite{pap2001}, \cite{golsar2003} \cite{ogilvie}, 
\cite{sargo2004}, 
\citet{dang}, \cite{kleydir}). This interest is driven
by the observed properties of the known extrasolar systems. One possible
origin of the non-zero eccentricities of extrasolar Jupiters is 
the planet-disc interaction, which is  most relevant in our calculations.
\cite{golsar2003} and \cite{sargo2004} have
estimated that eccentric Lindblad resonances can cause the eccentricity
growth if the planet can open a very clean gap in the disc.
\cite{pap2001} have found that for standard disc parameters
this mechanism of eccentricity growth can work for bodies on initially
circular orbits if the mass of the body 
is larger than 10 Jupiter masses. 
However, they do not exclude the possibility
that for
different disc parameters the mass of the planet in question might be smaller.
For our set of disc parameters ($h=0.03$ and $\nu =2\cdot 10^{-6}$)
the gap is wide enough and we can observe the
eccentricity growth.

The outcome of the scenario studied here  is that the Jupiter is
on a slightly
eccentric orbit and the Super-Earth is on an almost circular one.
This fact has interesting consequences from the astrobiological point
of view 
and we will come back to this issue in the last Section.

\subsection{The apsidal resonance}

\begin{figure}
\vskip 6.5cm
\includegraphics{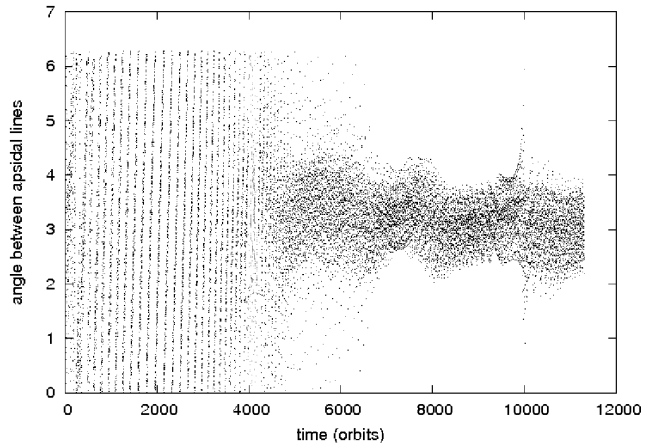}
\caption{\label{fig8}{The evolution of the angle between
the apsidal lines of the planetary orbits for model 4.
After the capture of the planet in the  apsidal resonance, the angle librates
around $\pi$.
}}
\end{figure}

Despite the absence of mean motion
commensurabilities between the Super-Earth and the Jupiter mass planet 
we have found the apsidal resonance in models 3, 4 and 5.
Both resonant angles
circulate in the whole range between  0 and 2$\pi$.  
At the beginning of the evolution the angle between the apsidal 
lines of the planetary orbits 
spans the whole interval between 0 and 2$\pi$ and then,
when the Super-Earth is already captured in the trap,
it starts to librate around the value of $\pi$  with an amplitude of about 
$90\,^{\circ}$, as it has been illustrated in Fig.~\ref{fig8}. 
This
behaviour
indicates
that the system is in apsidal resonance and in antialigned
configuration.   
A similar behaviour of the apsidal lines has been already observed in
extrasolar planetary systems, for example in $\upsilon And$ and
$47 UMa$ \citep{laughlin, barnes}.
It has been noticed \citep{barnes, barnes1} that planetary systems, as
for example 
$\upsilon And$ and
$47 UMa$, can display the secular behaviour close to the separatrix between
libration and circulation. The eccentricity growth can lead to the passage
from one mode of the secular behaviour to another. 
In our simulations the apsidal resonance occurs only in those models in
which the Super-Earth is trapped and the eccentricity of the Jupiter
increases. The eccentricity growth can be a promising mechanism for
establishing the apsidal resonance.
The full analysis of this issue requires simulations covering the
long timescales which are characteristic for the secular evolution.

\section{The evolution of the Super-Earth and the Sub-Jupiter }

\begin{figure*}
\begin{minipage}{160mm}
\vskip 6.5cm
\includegraphics{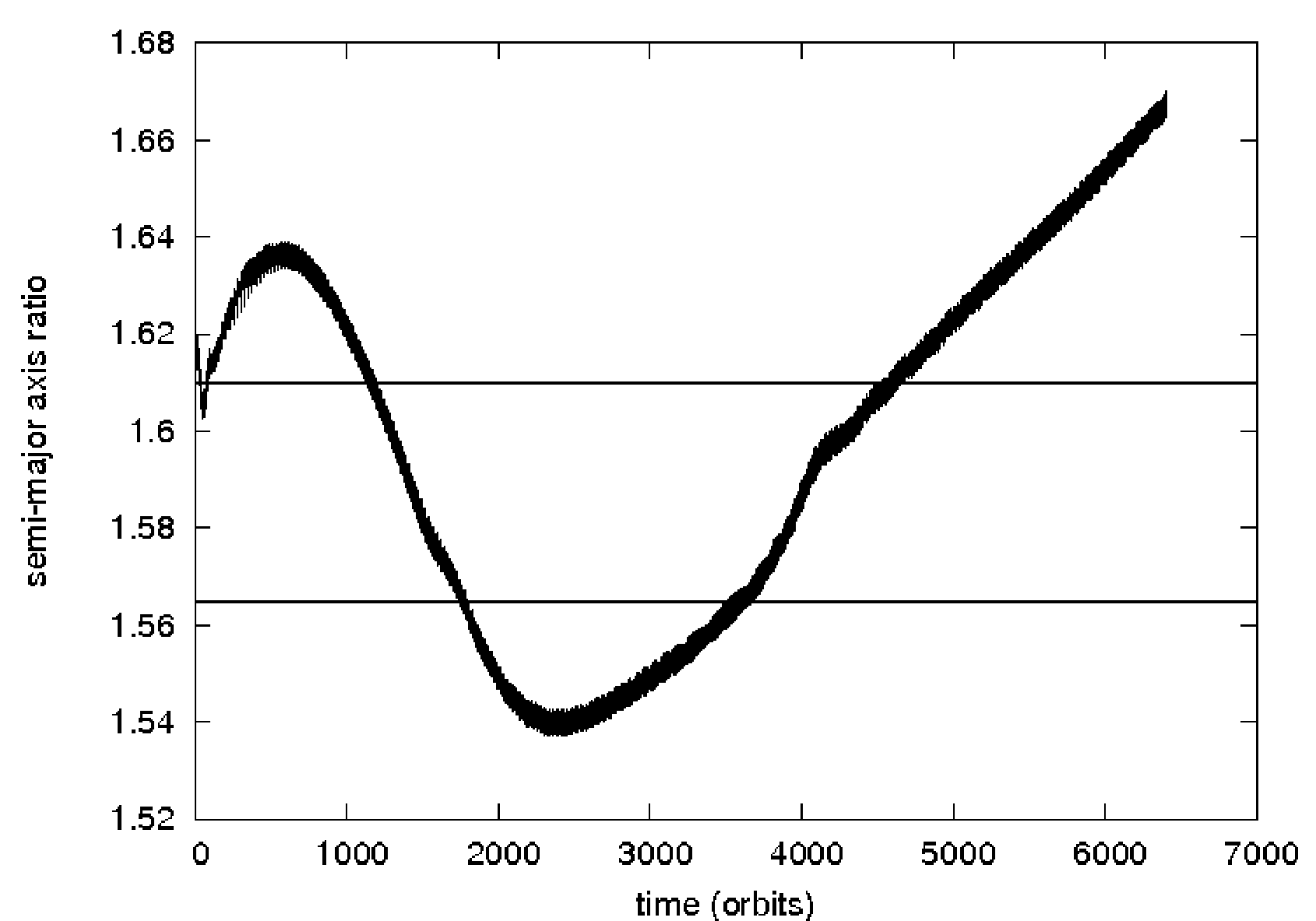}
\includegraphics{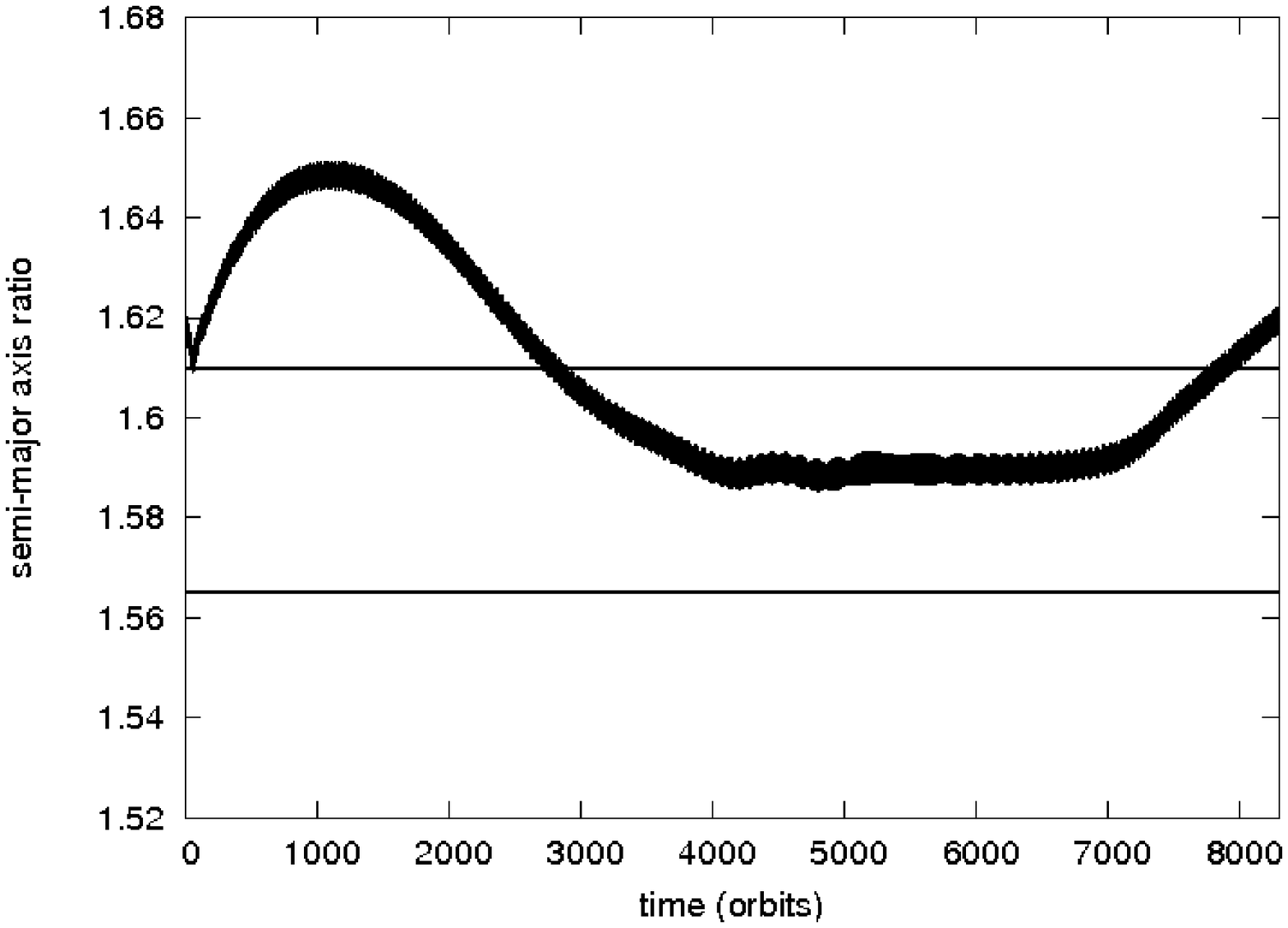}
\caption{\label{res}{The semi-major axis ratio of the planets for
model 12 (left panel) and model 14 (right panel). 
The horizontal lines denote the width of the 1:2 mean motion
resonance.
}}
\end{minipage}
\end{figure*}

As we have shown in Section~\ref{section:c}, the gap opened by the
Jupiter mass planet is too wide to attain 
any first order mean motion commensurability, so we have checked if 
resonances are
possible for Sub-Jupiter mass planets where the gap is narrower and
the 1:2 resonance is located outside the gap. 
In this way there is a chance that the Super-Earth is
captured into this commensurability before reaching the outer edge of
the gap where it can be trapped. 
We have tried the cases, in which the masses of the Sub-Jupiters are
0.7, 0.6 and 0.5 $M_J$. For the 0.7 $M_J$ case (model 11)
the gap is still too wide and the position of the 1:2 resonance is still 
inside it. In the two other cases the gap widths allow the possibility
of resonant 
locking. If the mass of the gas giant decreases, its migration rate
increases, because  such 
migrators cannot entirely open the gap and experience an additional torque
which pushes them faster toward the star 
\citep{edgar}. This means that it is more difficult to get convergent 
migration
in the case of Sub-Jupiters. On the contrary, for low-mass planets
the migration rate increases with increasing mass of the planet.
For this reason, in this Section we have taken the mass of the Super-Earth
to be equal to $10 M_\oplus$, instead of $5.5 M_\oplus$, in order to get
the fastest possible migration for the Super-Earth.
In a first set of experiments we have chosen for the kinematic
viscosity the value $\nu =2\cdot 10^{-6}$ and for the aspect ratio
 the typical
value $h=0.05$.
The parameter which has been allowed to vary in order to obtain
convergent migration is the surface density $\Sigma$.
Let us note that for
  $\Sigma$ higher than $2\Sigma_0$ we enter into the
regime of the runaway type III migration of the gas giant \citep{maspap}.
 In this case the Sub-Jupiter migrates very
fast and reaches the inner part of the disc after just a few hundreds
of orbits. 
In discs with lower surface densities the migration was always
divergent.
In a second run of experiments, we have fixed the surface density
to be $\Sigma =\Sigma_0$ trying instead
different aspect ratios.
For the Sub-Jupiter gas giant we have managed to achieve the
convergent migration of the planets  for $h=0.03$
(model 12). 
For this choice of
parameters, the relative migration of the planets is too fast 
and they passed
through the 1:2 mean motion resonance without being captured into this
commensurability. They migrated further
until the Super-Earth was trapped at the edge of
the gap after 2400 orbits. Successively the migration reversed as it is
shown in 
Fig. \ref{res} (left panel)
and the planets passed again through the resonance region without
being captured. 
In order to slow down the relative radial motion of the planets, we
have assumed in model 13 the
higher aspect ratio of the disc $h=0.04$, without 
changing other parameters.
In this case, the relative migration of the planets 
at the beginning was slowly convergent but later on became
divergent.

Finally, we have achieved the 1:2 mean motion resonance for a disc with 
$h=0.03$, $\nu=2 \cdot 10^{-6}$ and $\Sigma=0.5\Sigma_0$ (model
14). The lower surface 
density causes a slower migration of the Super-Earth and the
convergent relative migration of the planets was slower than in model
12. In Fig. \ref{res} (right panel)
we present the semi-major axis ratio of the planets. 
The divergent migration during the first 1000 orbits
is caused by the gas giant during the gap opening process.
It can be seen that
after 3000 orbits the planets approach the commensurability, 
both resonant angles and the angle between
apsidal lines starts to librate around $\pi$ and the eccentricities
of the planets 
increase. 
However, at the end, the planets migrate out of the resonance.
The Super-Earth  
with mass of 10 $M_{\oplus}$ opens a shallow
dip, which moves along with the migrating planet. 
The superposition of the dip and the outer edge of the gap creates a trap
and 
leads to the capture of the Super-Earth. 

From the above investigations we conclude that 
mean motion resonances between a gas giant and a Super-Earth which is
located further away in the disc should be rare. If the mass of the gas giant 
is higher
than $0.7M_J$, the gap opened by the planet is wide and the
Super-Earth is
captured in the trap before reaching the mean motion commensurability.
 For lower mass gas giants the gap is narrower and allows in
 principle the
occurrence of the 1:2 resonance, but the migration of the gas giant is too
fast for that. 
Our analysis, in which different disc parameters have been examined, shows
that it is
possible to obtain the short-lasting 1:2 mean motion commensurability 
between a Sub-Jupiter
mass gas giant on the internal orbit and a Super-Earth on the external
 one only for very thin discs with low surface density.

\section{Discussion and conclusions}

The large-scale orbital migration in young planetary systems might play an
important role in shaping up their architectures. The tidal gravitational
forces are able to rearrange the planet positions  according to their
masses and disc parameters. The final configuration after the disc
dispersal might 
be what we actually observe in extrasolar systems.
In particular, in a system with a gas giant and  a Super-Earth the 
convergent migration can lead to mean
motion resonant locking or the Super-Earth
is captured at the outer edge of the gap. The final outcome depends 
on the masses of the planets 
and on the disc parameters. 
We have investigated the evolution of the Super-Earth and of the gas giant 
embedded in a gaseous disc when the 
Super-Earth is on the external orbit and the gas giant on the internal
one 
similarly as in PN08. However, we have concentrated here 
on  systems with outer planet no more massive than $10 M_\oplus$ and we
have explored in details the evolution of such systems extending in
this way the
investigations done in PN08.
We have changed the initial radial location of the Super-Earth, the
masses of the
planets and the disc parameters in order
to check the  possibility of  locking into the 2:3 and 1:2
commensurabilities.
What we have found is that it is difficult to obtain mean motion resonant 
configurations between planets.
 For the set of disc parameters and planet masses considered in this
 paper, it turns out that either the migration is divergent, or
the Super-Earth migrates faster than the Jupiter until it is trapped at  the 
outer edge of the gap opened by the gas giant. Thus, the trapping of
the Super-Earth at the edge of the gap prevents the capture in
the mean motion resonance because the gap is too wide and the resonant
region is inside it.

In our studies we have investigated the behaviour of a Super-Earth,
which means a planet with a mass not bigger than $10M_{\oplus}$, in the
presence of a gas giant.
Switching on the accretion, would result in exceeding the mass limit 
for a Super-Earth, 
 so we didn't take accretion 
into account.  
However, if we allow the planets to accrete the matter than the planet 
is released 
from the
trap and it forms the resonant structure, as it has been shown in PN08. 

If the mass of the gas giant is low enough the gap is narrow
and the 1:2 mean motion resonance is allowed.
For a system with a Sub-Jupiter mass planet on the internal orbit and a
Super-Earth
on the external one, we have observed the resonant configuration only for
very thin discs with low surface density. 
The resonant configuration did not hold during the evolution, so 
no mean motion resonance has been observed at the end of our simulations.
Moreover, for a gas giant planet with Jupiter mass 
we have obtained a configuration in which
the apsidal resonance is present if the Super-Earth is captured at the edge
of the gap. Similar structures between apsidal lines
are  actually 
observed in $\upsilon And$ and
$47 UMa$ extrasolar planetary systems.

The disc properties used here have been determined by
the requirement of convergent migration in the disc.
In the relatively thin disc with aspect ratio  
$h=0.03$ (models 3-6) we have found that the eccentricity of the gas
giant increases, while the eccentricity of the Super-Earth remains 
very low. 
The growth of the Jupiter eccentricity is modest
but well pronounced. 

The fact that the Super-Earth remains on an almost circular
orbit can have important implications
for the habitability of such planets. As an example, let us consider a system
containing a gas giant inside the habitable zone. Then, small planets or
planetary embryos can survive the evolution of the system being captured at
the outer edge of the gap opened by the giant planet and thus can be located
exactly in the habitable zone. 
One of the possible places to look for
such configurations is the system  HD 27442. Another interesting
object is  $\rho$CrB, a star  with a Jupiter-like
planet which has the eccentricity equal to 0.04, like the system discussed in
this paper.
Future observations will be able to test the proposed
scenario, showing the way in which  planetary migration 
can shape the planetary orbits.

\section*{Acknowledgments}
This work has been partially supported by MNiSW grant N203 026 32/3831
(2007-2010) and  MNiSW PMN grant - ASTROSIM-PL ''Computational Astrophysics.
The formation  and evolution of structures in the universe: from planets to
galaxies'' (2008-2010).
The simulations reported here were performed using the
Polish National Cluster of Linux Systems (CLUSTERIX) and the HAL9000
cluster of the Faculty of Mathematics and Physics of the University of
Szczecin. We wanted to thank the referee for valuable comments which
helped to improve the manuscript.
We are grateful to John Papaloizou for enlightening discussions.
We wish also to thank Adam {\L}acny for his helpful
discussions and comments.
Finally, we are indebted to
Franco Ferrari for his continuous support in the development of our
computational 
techniques and computer facilities and for reading the manuscript.

\label{lastpage}

\end{document}